\documentclass[12pt,preprint]{aastex}

\slugcomment{Draft version \today}

\shorttitle{Rotation Measure due to IGMF}
\shortauthors{Akahori and Ryu}

\def\sfrac#1/#2{\kern.1em\raise.5ex\hbox{\the\scriptfont0 #1}\kern-.1em
/\kern-.15em\lower.25ex\hbox{\the\scriptfont0 #2}}

\begin{document}
\title{Faraday Rotation Measure due to the Intergalactic Magnetic Field II:
       the Cosmological Contribution}
\author{Takuya Akahori$^1$ and Dongsu Ryu$^2$\altaffilmark{,3}}
\affil{$^1$Research Institute of Basic Science, Chungnam National University,
Daejeon, Korea; akataku@canopus.cnu.ac.kr\\
$^2$Department of Astronomy and Space Science, Chungnam National
University, Daejeon, Korea; ryu@canopus.cnu.ac.kr}
\altaffiltext{3}{Author to whom any correspondence should be addressed.}

\begin{abstract}

We investigate the Faraday rotation measure (RM) due to the intergalactic magnetic field (IGMF) through the cosmic web up to cosmological distances, using a model IGMF based on turbulence dynamo in the large-scale structure of the universe. By stacking the IGMF and gas density data up to redshift $z=5$ and taking account of the redshift distribution of polarized background radio sources against which the RM is measured, we simulate the sky map of the RM. The contribution from galaxy clusters is subtracted from the map, based on several different criteria of X-ray brightness and temperature. Our findings are as follows. The distribution of RM for radio sources of different redshifts shows that the root-mean-square (rms) value increases with redshift and saturates for $z \ga 1$. The saturated value is RM$_{\rm rms} \approx$ several ${\rm rad~m^{-2}}$. The probability distribution function of $|{\rm RM}|$ follows the lognormal distribution. The power spectrum has a broad plateau over the angular scale of $\sim 1 - 0.1^\circ$ with a peak around $\sim 0.15^\circ$. The second-order structure function has a flat profile in the angular separation of $\ga 0.2^\circ$. Our results could provide useful insights for surveys to explore the IGMF with the Square Kilometer Array (SKA) and upcoming SKA pathfinders.

\end{abstract}


\section{Introduction}
\label{s1}

The cosmic web of filaments and clusters of galaxies, which is predicted in the highly successful $\Lambda$CDM cosmology \citep{bkp96}, is filled with ionized plasma, {\it i.e.}, the intergalactic medium (IGM) \citep{co99,krcs05}. The hot gas with $T > 10^7$ K is found mostly in the intracluster medium (ICM) and cluster outskirts, and the gas with $10^5$ K $ < T < 10^7$ K, which is also referred as the Warm Hot Intergalactic Medium (WHIM), is distributed mostly in filaments. They were heated mostly by cosmological shock waves which were formed in the course of the large-scale structure (LSS) formation of the universe \citep{rkhj03,psej06,krco07,skillman08,hoeft08,vazza09}. The diffuse gas with $T < 10^5$ K resides mainly in sheetlike structures and voids.

The IGM is expected to be permeated by magnetic fields just as the interstellar medium within galaxies. Theoretical studies have predicted the existence of the intergalactic magnetic field (IGMF). A number of mechanisms that can create seed magnetic fields have been suggested; besides processes based on inflation and phase transitions \citep[see, e.g.,][for review]{wrss10} and plasma physical processes \citep[see, e.g.,][for review]{rstt10} in the early universe, astrophysical processes include the field generations during the reionization of the universe \citep{gfz00,lap05}, by first stars \citep{xu08,ads10}, and at cosmological shock waves \citep{kcor97,rkb98}, as well as the leakage of magnetic fields and cosmic ray particles from galaxies \citep{ddlm09,mb11}. The seed fields can be further amplified by flow motions induced by the hierarchical clustering during the LSS formation \citep{kcor97,rkcd08,sbsa10}.

Faraday rotation, the rotation of the plane of linearly-polarized radio emission due to the birefringence of magneto-ionic media, provides an observational means to exploring the IGMF. There have been a number of studies of the intracluster magnetic field (ICMF) through observations of rotation measure (RM) \citep[see][for a review]{ct02}. For instance, RMs of hundreds ${\rm rad\ m^{-2}}$ were observed in clusters, indicating an average strength of the ICMF to be $\sim 1$--10\ $\mu$G \citep{ckb01,cla04,gov10}. In addition, RM maps of clusters were analyzed to get the power spectrum of turbulent magnetic fields in the ICM; for instance, a Kolmogorov-like spectrum with a bending at a few kpc scale was found in the cooled core region of the Hydra cluster \citep{ve05}, and spectra consistent with the Kolmogorov spectrum were reported in the wider ICM for the Abell 2382 cluster \citep{gmgp08} and for the Coma cluster \citep{bfmg10}.

Studies of RM outside clusters, through the cosmic web, are, on the contrary, still scarce \citep[e.g.,][]{xkhd06}, because detecting small RM is difficult with current observational facilities, and also removing the galactic foreground is not a trivial task. Nevertheless, recently, constraints for the RM through the LSS have been discussed by several authors \citep{tss09,mao10,sch10,sts11}. For instance, \citet{sch10} argued that in the catalog of \citet{tss09}, the extragalactic contribution to the width of RM distribution would be $\sigma_{\rm RM,EG}\sim 6$~${\rm rad~m^{-2}}$. But the measurement error is still large, $\sigma_{\rm errRM}\sim$10~${\rm rad~m^{-2}}$; that is, so far, the nature and origin of the IGMF outside clusters has not been well constrained with RM studies. However, the next generation radio interferometers including the Square Kilometer Array (SKA), and upcoming SKA pathfinders, the Australian SKA Pathfinder (ASKAP) and the South African Karoo Array Telescope (MeerKAT), as well the Low Frequency Array (LOFAR) will enable us to investigate the IGMF outside clusters with high-sensitivity RM observations \citep[see, e.g.,][]{cr04,beck09,kra09,glt10}.

Theoretical predictions for the RM due to the IGMF have been made with model IGMFs and simulations of the LSS formation \citep{rkb98,dub08,cr09,ds09,sndbd10,ar10}. Especially, based on a physically motivated model, in which a part of the gravitational energy released during the LSS formation is transferred to the magnetic field energy as a result of the turbulent amplification of weak seed fields, \citet{rkcd08} (R08) proposed that the IGMF follows largely the matter distribution in the cosmic web and the mean strength would be $\langle B\rangle \sim 10$ nG in filaments in the local universe at redshift $z=0$. Studying various characteristic length scales of magnetic fields in MHD turbulence simulations, \citet{cr09} (CR09) proposed that the RM coherence length of the IGMF in filaments would be a few $\times\ 100\ h^{-1}$ kpc. Based on the model IGMF of R08, CR09 predicted that the-root-mean-square (rms) value of RM through a single filament would be of order $\sim 1\ {\rm rad\ m^{-2}}$. And using the same model IGMF, we simulated RM in the local universe \citep{ar10} (AR10). We found that with the path length larger than the coherence length of the IGMF, the inducement of RM is a random walk process, but the resultant RM is dominantly contributed by the density peaks along the line of sight (LOS). The rms value of RM through a single filament was estimated to be $\sim 1\ {\rm rad\ m^{-2}}$, which is in agreement with the prediction by CR09. We also found that the probability distribution function (PDF) of $|{\rm RM}|$ follows the log-normal distribution, and the power spectrum of the RM peaks at a scale of order $\sim 1\ h^{-1}$ Mpc. 

In this paper, we extend the RM study of AR10 in the present-day, local universe by including the cosmological contribution. By taking account of the redshift evolution of the LSS and the IGMF as well as the redshift distribution of radio sources against which RM is measured, we calculate the RM through the cosmic web up to $z=5$. We then examine the spatial and redshift distributions of the RM and their statistics. In Sections 2 and 3, we describe our models results, respectively. Discussion is in Section 4, and summary and conclusion follow in Section 5.

\section{Models}
\label{s2}

\subsection{Large-Scale Structure of the Universe}
\label{s2.1}

For the LSS of the universe, we used structure formation simulations of a $\Lambda$CDM universe with the following values of cosmological parameters: $\Omega_{\rm b0}=0.043$, $\Omega_{\rm m0}=0.27$, $\Omega_{\rm \Lambda 0}=0.73$, $h\equiv H_0/(100~{\rm km/s/Mpc})=0.7$, $n=1$, and $\sigma_8=0.8$. The simulations were performed using a particle-mesh/Eulerian, cosmological hydrodynamic code \citep{rokc93}. A cubic region of comoving volume $(100\ h^{-1}{\rm Mpc})^3$ was reproduced with $512^3$ uniform grid zones for gas and gravity and $256^3$ particles for dark matter, so the spatial resolution is $195\ h^{-1}$ kpc. Sixteen simulations with different realizations of initial condition were used to compensate cosmic variance. They are the same set of simulations used in AR10.

\subsection{Intergalactic Magnetic Field}
\label{s2.2}

As in AR10, we employed the model described in R08. It assumes that turbulent-flow motions are induced via the cascade of the vorticity generated due to cosmological shocks during the formation of LSS, and the IGMF is produced as a consequence of the amplification of weak seed fields of any origin through the stretching of field lines by the flow motions. Then, it can be modeled that a fraction of the turbulent-flow energy, $\varepsilon_{\rm turb}$, is converted to the magnetic energy, $\varepsilon_B$, as
\begin{equation}\label{eq1}
\varepsilon_B = \phi \left(\frac{t}{t_{\rm eddy}}\right) \varepsilon_{\rm turb},
\end{equation}
where the conversion factor, $\phi$, depends only on the eddy turnover number, $t/t_{\rm eddy}$. Here, the eddy turnover time is defined as the reciprocal of the vorticity at driving scales, $t_{\rm eddy} \equiv 1/\omega_{\rm driving}$ (${\vec \omega} \equiv {\vec \nabla}\times{\vec v}$). The local vorticity and turbulent-flow energy density were calculated from the data of the structure formation simulations described above, and the age of the universe at the redshift $z$ was used for $t$. The functional form for the conversion factor was derived from a separate, incompressible, magnetohydrodynamic (MHD) simulation of turbulence dynamo (R08). Then, the magnetic energy density was calculated according to Equation \ref{eq1}, and the strength of the IGMF as $B = \sqrt{8\pi\varepsilon_B}$. The average strength of the resulting model IGMF for the WHIM is $\langle B\rangle \sim 10$ nG or $\langle \rho B\rangle/\langle \rho\rangle \sim 0.1\ \mu$G at $z=0$. For the direction of the IGMF, we used that of the passive fields from structure formation simulations, in which weak seed fields were generated through the Biermann battery mechanism \citep[][]{bie50} at cosmological shocks and evolved passively, ignoring the back-reaction, along with flow motions \citep{kcor97,rkb98}.

The validity of our model IGMF was discussed in details in AR10. It was argued that our model IGMF would produce reasonable results for the RM through the cosmic web in the local universe. The simulated passive fields reproduce the directions that show the expected correlation with those of vorticity. The RM coherence length for the IGMF in filaments is estimated to be a few to several $\times 100$ kpc at $z=0$, which agrees with the estimation of CR09. It is a few times larger than the grid resolution of our simulations, $195\ h^{-1}$ kpc. On the other hand, the coherence length for the IGMF in clusters is expected to be a few $\times 10$ kpc (CR09), which is smaller than the grid resolution. So the application of our model IGMF to clusters would results in an erroneous estimation of RM. However, here we study the RM through the cosmic web outside clusters, and so we will remove the contribution from clusters to RM (see Section 2.8).

Our model predicts that the IGMF for the WHIM was a bit stronger in the past; for instance, $\langle B\rangle \sim 30$ nG for the gas with $10^5$ K $< T <$ $10^7$ K at $z=5$ (see Figure 2 of R08). It is because the density of the WHIM was higher in the past, although the magnitudes of vorticity and the vortical component of velocity were smaller. Our model also predicts stronger IGMF for the hot gas with $T > 10^7$ K in the past. However, the IGMF averaged over the entire computational volume was weaker in the past, because the volume and mass fractions of strong-field regions were smaller.

\subsection{Data Stacking}
\label{s2.3}

To reproduce the cosmic space, we stacked simulation boxes up to $z=5$, basically following the conventional manner of cosmological data stacking \citep[e.g.,][]{sblt00}. We took outputs at $z_{\rm out}=$ 0, 0.2, 0.5, 1.0, 1.5, 2.0, 3.0, and 5.0, and put the outputs of the closest redshift for stacking boxes; 56 boxes were required to reach $z=5$. The stacking boxes were randomly selected from sixteen simulations and randomly rotated to avoid any artificial coherent structure along the LOS. If a LOS went out of the boundary of a box, we applied the periodic boundary condition in which we replicated the box across the LOS. We also carried out the calculation with the open boundary condition in which we did not replicate the box but put different one from sixteen simulation outputs, and confirmed that the statistical properties of RM we examined are not sensitive to the boundary conditions.

\subsection{Observer Locations}
\label{s2.4}

Our Galaxy is located in the Local Group. Since the Local Group is probably filled with the magnetized WHIM, the observed RM could contain a contribution from the IGMF of the Local Group (see Section 3.1). It would have been ideal to place the ``observer'' where the IGMF and gas density are similar to those in the Local Group. Unfortunately, however, not much is known about the physical states of the Local Group. Hence, following \citet{dkrc08}, instead, we selected groups of galaxies identified in the simulation data that have similar halo gas temperatures to the Local Group. We considered two cases: groups at $z=0$ with 0.05 keV $<$ $k T_X$ $<$ 0.15 keV and groups at $z=0$ with 0.05 keV $<$ $k T_X$ $<$ 0.5 keV \citep{rp01}. Here, $T_X$ is the X-ray temperature (see Section 2.7). Assuming that these groups are located in a physical environment similar to that of the Local Group, we placed ``mock observers'' at the center of the groups.

\subsection{Survey Setup}
\label{s2.5}

We aim to simulate wide-field, high-sensitivity RM surveys that will be achieved with future radio interferometers. For instance, the upcoming POSSUM (Polarization Sky Survey of the Universe's Magnetism) project, one of the major surveys planned for the ASKAP, will produce an RM map that will be a dramatic improvement over existing maps. The RM map from the POSSUM will have a field-of-view (FOV) of $\theta^2 \simeq 30$ ${\rm deg^2}$ and an average source separation of $\left<\Delta\theta\right> \simeq 0.1^\circ$ ($\sim 6$ arcmin) corresponding to an RM grid of $\sim 100$ ${\rm RM~deg^{-2}}$ \citep[see, e.g.,][]{glt10}. The SKA will survey $\sim 10^7$ polarized radio sources over the whole sky with $\left<\Delta\theta\right> \simeq 1$ arcmin \citep[see, e.g.,][]{beck09}. Intending to produce a simulated map of RM, we adopted $\theta^2 = 200$ ${\rm deg^2}$ $= (14.14)^2$ ${\rm deg}^2$ and $\Delta\theta = \sqrt{200\ {\rm deg^2 / 2048^2}} = 0.414$ arcmin with one source in each of $2048^2$ equally spaced pixels. Specific simulations for the SKA survey and the ASKAP POSSUM that consider the feasibility of observations such as the effects of the measurement error, coarser spacing, randomly placed sources, intrinsic RM, and so on, as well as the galactic foreground will be discussed in separate papers.

\begin{figure}[t]
\figurenum{1}
\epsscale{0.8}
\plotone{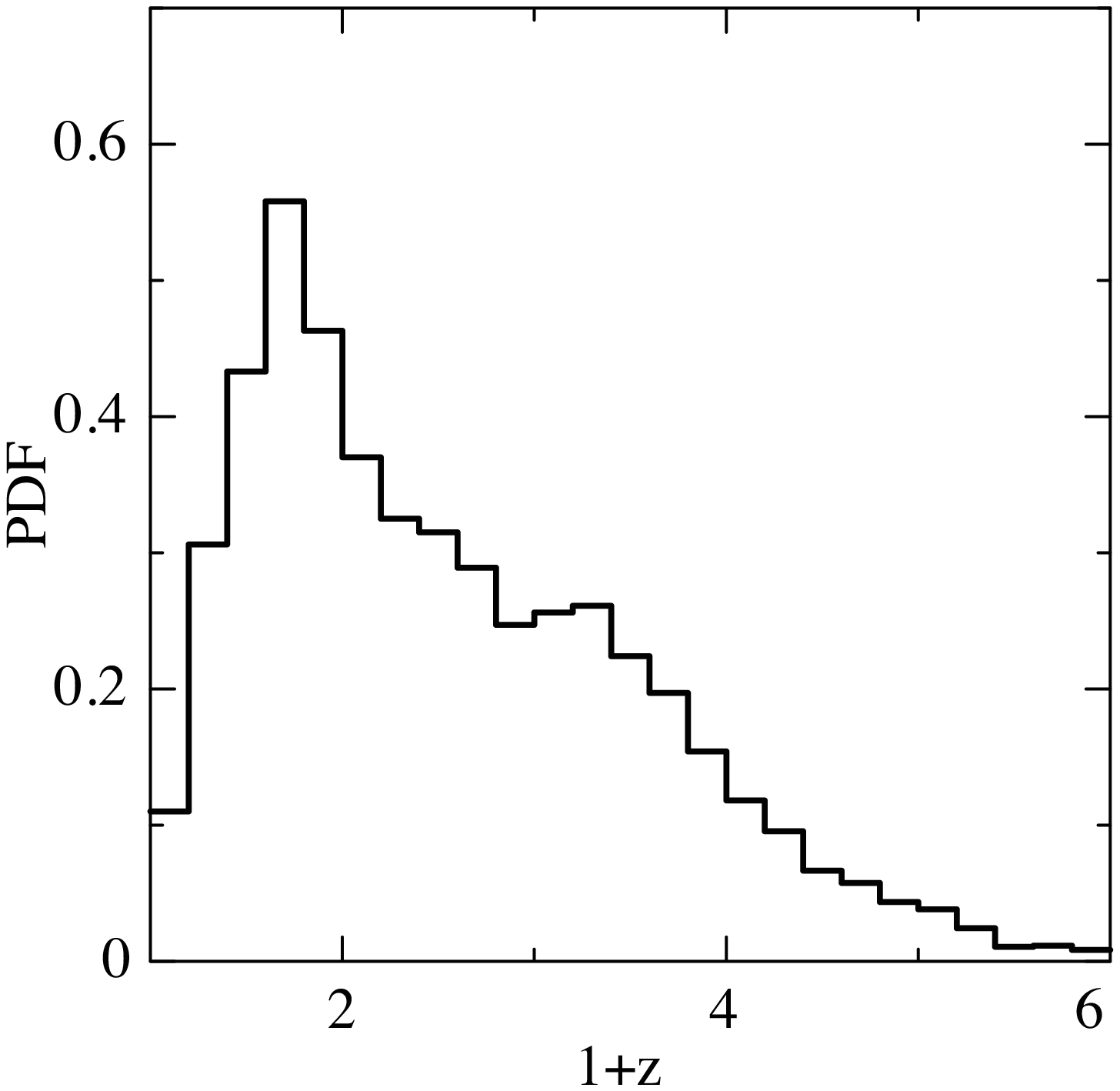}
\caption{Redshift probability distribution function of radio sources based on the expected number of observable FR I + FR II galaxies by the SKA \citep{will08}.\label{f1}}
\end{figure}

For the redshift distribution of radio sources, we employed a distribution based on the estimation of the detectable radio galaxies by the SKA and ASKAP \citep[][see also Figure \ref{f1}]{will08}, in which the observed FR I and FR II galaxies are taken into consideration. The number of available sources drops substantially at $z \ga 5$, and that is the reason why the data stacking was made up to $z=5$ (see Section 2.3).

\subsection{Faraday Rotation Measure}
\label{s2.6}

RM is a measure of the Faraday rotation of polarized radio emission against a background source, defined by
\begin{equation}\label{eq2}
{\rm RM}\equiv \frac{\psi}{\lambda_{\rm obs}^2} = \frac{e^3}{2\pi m_{\rm e}^2 c^4} 
\int_{0}^{l_{\rm s}} n_{\rm e}B_\parallel \frac{\lambda(l)^2}{\lambda_{\rm obs}^2}dl,
\end{equation}
where $\psi$ is the rotated angle of the plane of polarized radio emission, $l_{\rm s}$ is the path length up to the source, $n_{\rm e}$ is the thermal electron density, $B_\parallel$ is the LOS component of the magnetic field, $\lambda_{\rm obs}$ is the observed wavelength, and $\lambda(l)$ is the wavelength along the path length of polarized radio emission \citep[e.g.,][]{rl79}. For a source at cosmological distance of redshift $z_s$, it is modified to
\begin{equation}\label{eq3}
{\rm RM}= \frac{e^3}{2\pi m_{\rm e}^2 c^4} 
\int_{0}^{l_{\rm s}(z_{\rm s})}
(1+z)^{-2} n_{\rm e}(z) B_\parallel(z) dl(z),
\end{equation}
where $n_{\rm e}(z)$, $B_\parallel(z)$, and $l(z)$ are the quantities in proper coordinates.
The path length is given as
\begin{equation}\label{eq4}
dl(z) = \frac{c\ dz}{H_0(1+z)\sqrt{\Omega_{\rm m0}(1+z)^3+\Omega_{\rm \Lambda 0}}},
\end{equation}
for the flat universe with $\Omega_{m0} + \Omega_{\Lambda0} = 1$ \citep[e.g.,][]{peeb93}.

We calculated Equation (\ref{eq3}) by counting the contribution from computational grid zones along stacked boxes up to mock sources as
\begin{equation}\label{eq5}
{\rm RM}~({\rm rad~m^{-2}}) = 8.12\times 10^5 \sum_{i=1}^{N_s(z_s)} (1+z_i)^{-2}
\cdot n_{\rm e}(z_i) \cdot  B_\parallel (z_i) \cdot \Delta l (z_i),
\end{equation}
where $N_s(z_s)$ is the number of grid zones up to sources at $z_s$. The density and magnetic field strength are in units of ${\rm cm^{-3}}$ and $\mu$G, respectively. For $\Delta l(z)$, we used the proper size of grid zone, which is $0.195~h^{-1}(1+z)^{-1}$ Mpc. With the simulation FOV of $\theta^2 = (14.14)^2$ ${\rm deg}^2$, the maximum tilt of LOSs relative to the coordinates of simulation box is $\cos(\theta/2) = 0.99$, so we ignored the effect on $\Delta l(z)$.

\subsection{X-ray Emission}
\label{s2.7}

In the hot gas with $T \ga 10^7$ K, X-ray emission is produced mainly by thermal bremsstrah\-lung. We hence considered only the bremsstrah\-lung emission from electrons, and neglected line emissions from ions. The emissivity of thermal bremsstrahlung can be expressed as 
\begin{equation}\label{eq6}
\varepsilon^{\rm ff}~({\rm erg~s^{-1}~cm^{-3}})
=7.77\times 10^{-38}T^{-1/2} n_e^2
\cdot \int_{\nu_1}^{\nu_2} \bar{g}(T,\nu) \exp\left(-\frac{h_{\rm p}\nu}{k_{\rm B}T}\right) d\nu,
\end{equation}
where $T$, $n_e$, and $\nu$ are in units of K, ${\rm cm^{-3}}$, and Hz, respectively, and $h_{\rm p}$ and $k_{\rm B}$ are the Planck and Boltzmann constants, respectively  \citep{rl79}. We adopted the approximate Gaunt factor, $\bar{g} \approx 0.9(h_{\rm p}\nu/k_{\rm B}T)^{-0.3}$, and used the bolometric emissivity, that is, $\nu_1 = 0$ and $\nu_2 = \infty$.

For the X-ray temperature of clusters and groups, we calculated the X-ray emissivity-weighted temperature as
\begin{equation}\label{eq7}
T_X=\int \varepsilon^{\rm ff} T dV\ \Bigg/\ \int \varepsilon^{\rm ff} dV,
\end{equation}
over a spherical volume of comoving radius $0.5 h^{-1}$ Mpc.

For the X-ray surface brightness and surface temperature up to redshift $z$, we first sought all the grid zones, $k$, and their proper volume, $V_k$, which enters within the angular beam of $(\Delta\theta)^2$. Note that $\Delta\theta = 0.414$ arcmin corresponds to the comoving size of a grid zone $195\ h^{-1}$ kpc at $z \simeq 0.6$; hence perpendicular to a LOS, less then one zone enters at lower redshifts, but more than one at higher redshifts within $(\Delta\theta)^2$. For each grid at redshift $z_k$, the X-ray luminosity was calculated as
\begin{equation}\label{eq8}
L_k = \int_{V_k} \varepsilon^{\rm ff} dV.
\end{equation}
Then, the X-ray surface brightness was calculated as
\begin{equation}\label{eq9}
S_X^* = \frac{1}{(\Delta \theta)^2} \sum_k^{{\rm up\ to\ }z}
\frac{L_k}{4\pi d_c(z_k)^2(1+z_k)^2},
\end{equation}
and the X-ray emissivity-weighted surface temperature was calculated as 
\begin{equation}\label{eq10}
T_X^* = \frac{1}{S_X^*(\Delta \theta)^2} \sum_k^{{\rm up\ to\ }z}
\frac{T_{Xk} L_k}{4\pi d_c(z_k)^2(1+z_k)^2},
\end{equation}
where $T_{Xk}$ is the temperature of grid zone $k$. Here, $d_c(z_k)$ is the comoving distance up to $z_k$, which is given as
\begin{equation}\label{eq11}
d_c(z_k) = \frac{c}{H_0} \int_0^{z_k} \frac{dz}{\sqrt{\Omega_{\rm m0}(1+z)^3+\Omega_{\rm \Lambda 0}}}.
\end{equation}

\subsection{Subtraction of Cluster Contribution}
\label{s2.8}

The contribution from clusters was subtracted from calculated RM based on different criteria using X-ray brightness and temperature. The simplest approach would be to exclude the volume associated with clusters in the cosmic space. We employed the following two criteria. In the first model, labeled ``TM7'', all the hot gas with $T > 10^7$ K was excluded. In the second model, labeled ``CLS'', the spherical region of comoving radius $1 h^{-1}$ Mpc around the clusters and groups with $T_X > 2$ keV  was excluded.

For simulations of RM surveys, however, a better approach would be the exclusion of the pixels that include clusters in the projected sky map. We would like to subtract the contribution from all clusters, but it is still hard to detect faint X-ray clusters. So we tried the following criteria. In the model labeled ``TS8'', all the pixels with $T_X^* > 10^7$ K and $S_X^* > 10^{-8}$ ${\rm erg~s^{-1}~cm^{-2}~sr^{-1}}$ were excluded; $10^{-8}$ ${\rm erg~s^{-1}~cm^{-2}~sr^{-1}}$ is close to the detection limit of current X-ray facilities \citep[e.g.,][]{hos10}. And in another model labeled ``TS0'', the pixels with $T_X^* > 10^7$ K and $S_X^* > 10^{-10}$ ${\rm erg~s^{-1}~cm^{-2}~sr^{-1}}$ were excluded; $10^{-10}$ ${\rm erg~s^{-1}~cm^{-2}~sr^{-1}}$ is intended to mimic the improved detection limit of future X-ray facilities. Here, $S_X^*$ and $T_X^*$ up to $z=5$ were used. Finally, we also considered the case where no volume or pixel was excluded, labeled ``ALL", for comparison.

\section{Results}
\label{s3}

\subsection{Contribution from Groups around Observers}
\label{s3.1}

\begin{figure}[t]
\figurenum{2}
\epsscale{0.8}
\plotone{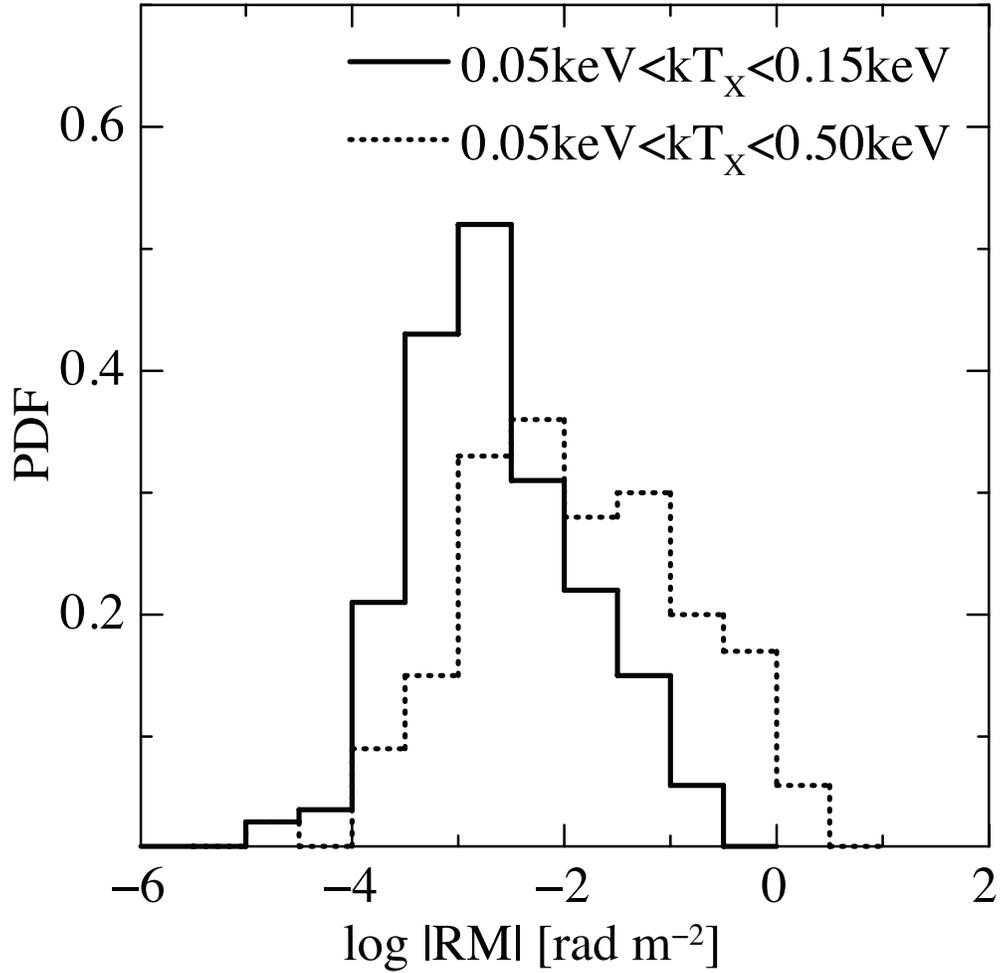}
\caption{PDF of $|{\rm RM}|$, the absolute value of RM, due to the local IGMF in groups where observers are located. Solid and dotted lines show the results with the temperature criteria for the groups, 0.05 keV $<$ $kT_X$ $<$ 0.15 keV and 0.05 keV $<$ $kT_X$ $<$ 0.5 keV, respectively. 200 simulations are carried out for each case.\label{f2}}
\end{figure}

We first examine the contribution to RM from the local IGMF around observers. For the two temperature range of groups containing observers, 0.05 keV $<$ $kT_X$ $<$ 0.15 and 0.05 keV $<$ $kT_X$ $<$ 0.5 keV (see Section 2.4), we set up 200 mock observers and calculated RM by integrating along LOSs up to $\sim 0.5$ $h^{-1}$ Mpc (2$\sfrac1/2$ grid zones) from observers. Figure \ref{f2} plots the PDF of $|{\rm RM}|$, the absolute value of resulting RM. The PDF peaks around $\sim 2 \times 10^{-3}\ {\rm rad~m^{-2}}$ and $\sim 6 \times 10^{-3}\ {\rm rad~m^{-2}}$ for 0.05 keV $<$ $kT_X$ $<$ 0.15 keV and 0.05 keV $<$ $kT_X$ $<$ 0.5 keV, respectively. The rms values are $\sim 5 \times 10^{-2}\ {\rm rad~m^{-2}}$ and $\sim 6 \times 10^{-1}\ {\rm rad~m^{-2}}$, respectively. The peak values can be estimated roughly as follows. Our model IGMF predicts $B_{\rm peak} \sim 10^{-3}\ \mu$G for groups \citep[see][]{dkrc08}. With $n_e \sim 10^{-5} - 10^{-4}$ cm$^{-3}$ for groups, and the coherence length of magnetic fields comparable to the size of grid zones, {\it i.e.,} $l_{\rm coherence} \sim 195\ h^{-1}$ kpc, we get ${\rm RM}_{\rm peak} \sim 10^{-3} - 10^{-2}$ ${\rm rad~m^{-2}}$. This may indicate that the contribution from the local IGMF is affected by the grid resolution. But as we will see in Section 3.3, this contribution makes only a small, negligible fraction of the RM through the cosmic web. Below we show only the results with 0.05 keV $<$ $kT_X$ $<$ 0.15; those with 0.05 keV $<$ $kT_X$ $<$ 0.5 keV are statistically indistinguishable.

\subsection{Two-Dimensional Map}
\label{s3.2}

We produced the simulated sky map of RM (using Equation (\ref{eq5})) with different cluster subtraction models, as well as the accompanying map of $S_X^*$ (using Equation (\ref{eq9})) and $T_X^*$ (using Equation (\ref{eq10})), for 200 different mock observers with different stacks, each of which has $2048^2$ pixels in a FOV of 200 ${\rm deg^2}$, as described in Section 2.5. The statistics below were obtained with the maps. In this section, we show maps for a typical case.

\begin{figure}[t]
\figurenum{3}
\epsscale{1.0}
\plotone{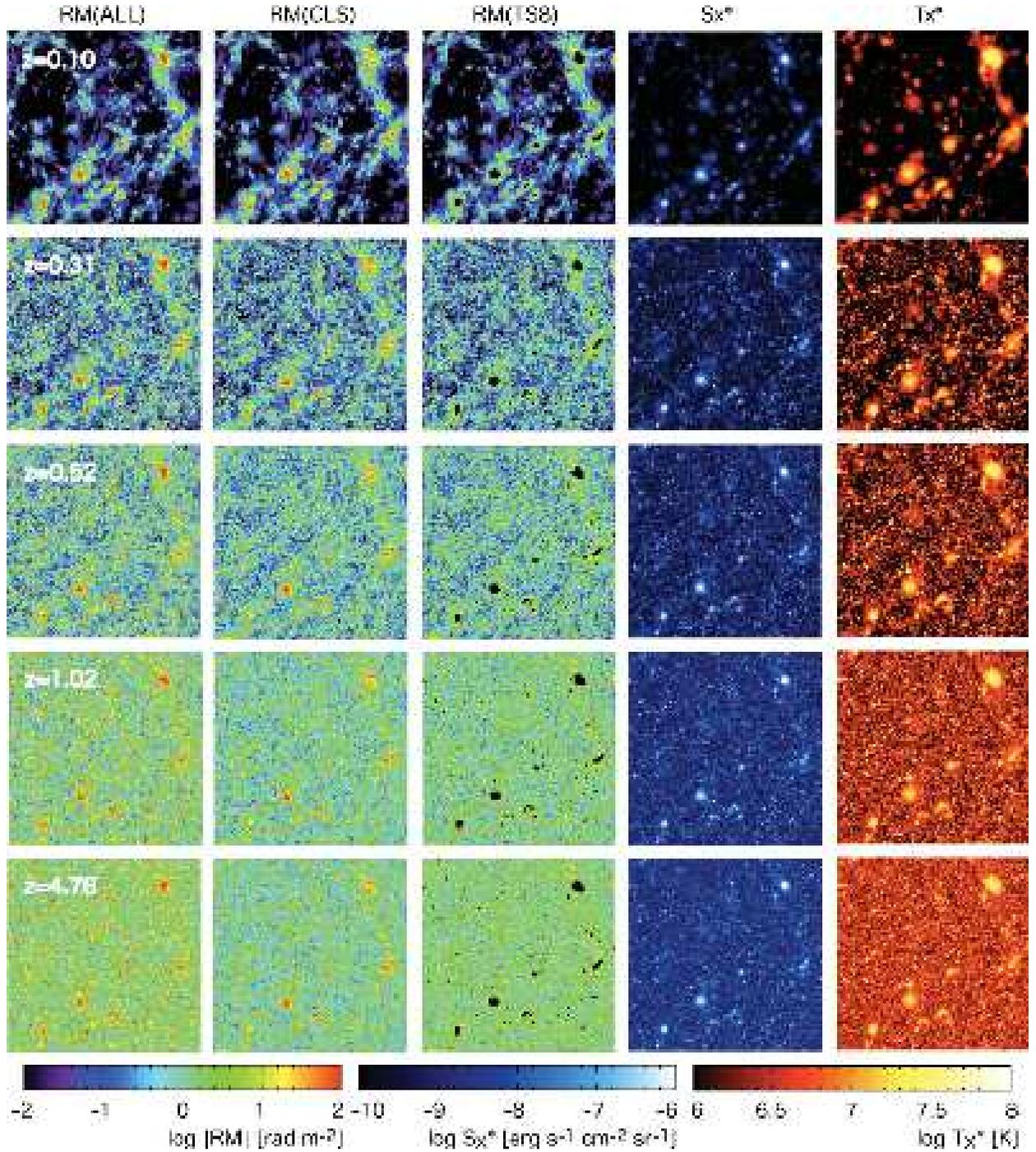}
\caption{Maps of RM for ALL, CLS, and TS8 models, X-ray surface brightness, $S_X^*$, and X-ray emissivity-weighted average temperature, $T_X^*$, integrated from $z=0$ up to the redshift depth shown. The angular size of each map is $14.14 \times 14.14$ deg$^2$.\label{f3}}
\end{figure}

Figure \ref{f3} shows RM maps for ALL and cluster subtraction models CLS and TS8, and $S_X^*$ and $T_X^*$ maps. To see the dependence on the redshift depth, the maps integrated up to five different epochs are presented.  With TS8, about $\sim 15 \%$ of pixels were subtracted. At the lowest redshift depth, X-ray bright sources, which can be identified as clusters or groups, are apparent. They have the angular size of up to a couple of degrees and the peak value of $|{\rm RM}| \ga 100~{\rm rad~m^{-2}}$. With CLS, some, but not all, contribution to RM from those clusters and groups was excluded. With TS8, most contribution from clusters and groups was removed. On the other hand, the contribution from filamentary structures was kept; filamentary structures with $|{\rm RM}| \sim 0.1\ -$ a few ${\rm rad~m^{-2}}$ are seen in CLS and TS8 as well as in ALL. At higher redshift depths, the value of $|{\rm RM}|$ is larger. This is attributed to the fact that the inducement of RM is a random walk process (AR10); the value increases with the increasing path length. Also at higher redshift depths, the angular size of RM structures is smaller, since the distance is larger and the proper size of filaments is smaller. At the depth of $z=5$, most of nearby filamentary structures are smeared out, and the sky distribution of RM appears close to be random.

\subsection{PDF and Root Mean Square}
\label{s3.3}

\begin{figure}[t]
\figurenum{4}
\epsscale{0.6}
\plotone{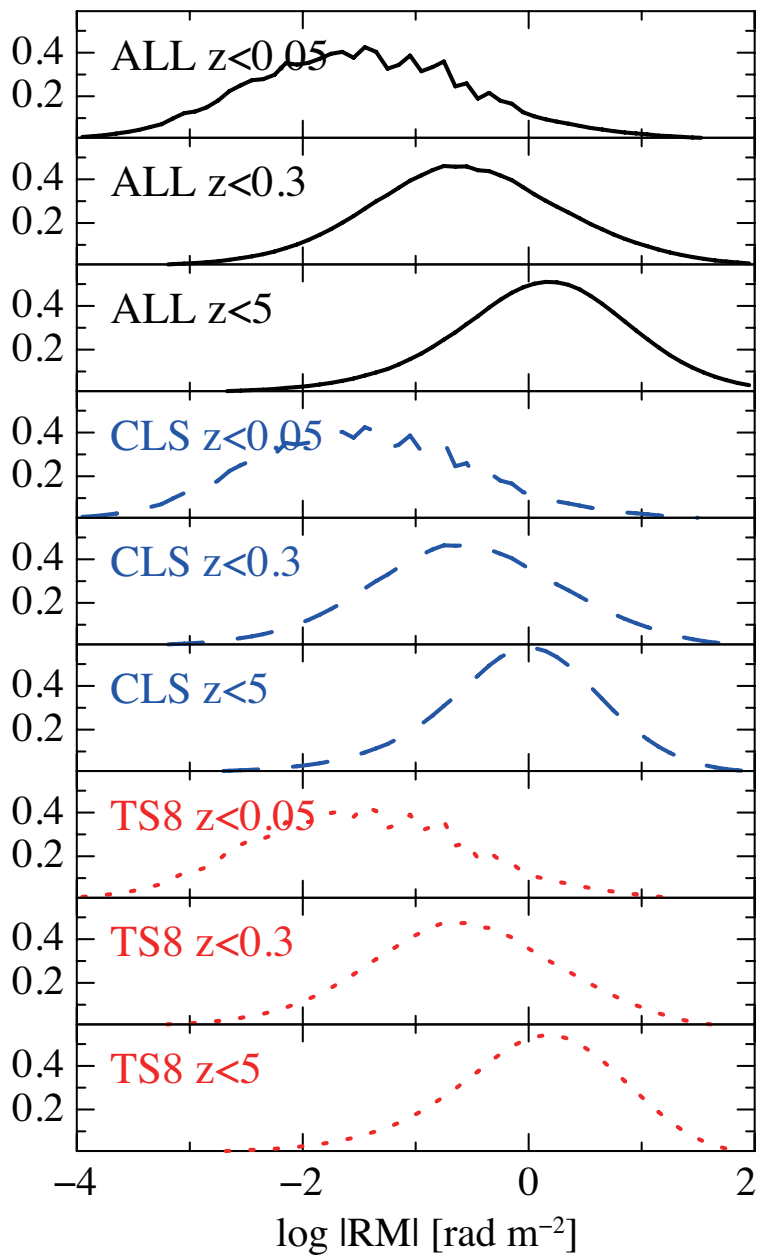}
\caption{PDF of $|{\rm RM}|$ integrated up to the redshift depth shown. The average for $2048^2 \times 200$ pixels is shown. Solid, dashed, and dotted lines show the results for ALL, CLS, and TS8 models, respectively.\label{f4}}
\end{figure}

We quantified the RM in $2048^2$ pixels of 200 maps, so in total $2048^2 \times 200$ pixels, first with the PDF and rms value. Figure \ref{f4} shows the PDF of $|{\rm RM}|$ for three different redshift depths, $z=0.05$, 0.3, and 5, for ALL, CLS, and TS8. As for the $|{\rm RM}|$ of the local universe (AR10), the PDF follows the lognormal distribution. For TS8, the peak value shifts from a few $\times 10^{-2}$ ${\rm rad~m^{-2}}$ at the depth of $z = 0.05$ to $\sim 1\ {\rm rad~m^{-2}}$ at $z=5$. Compared to CLS and TS8, ALL has a high-value tail due to the contribution from clusters and groups.

\begin{figure}[t]
\figurenum{5}
\epsscale{0.8}
\plotone{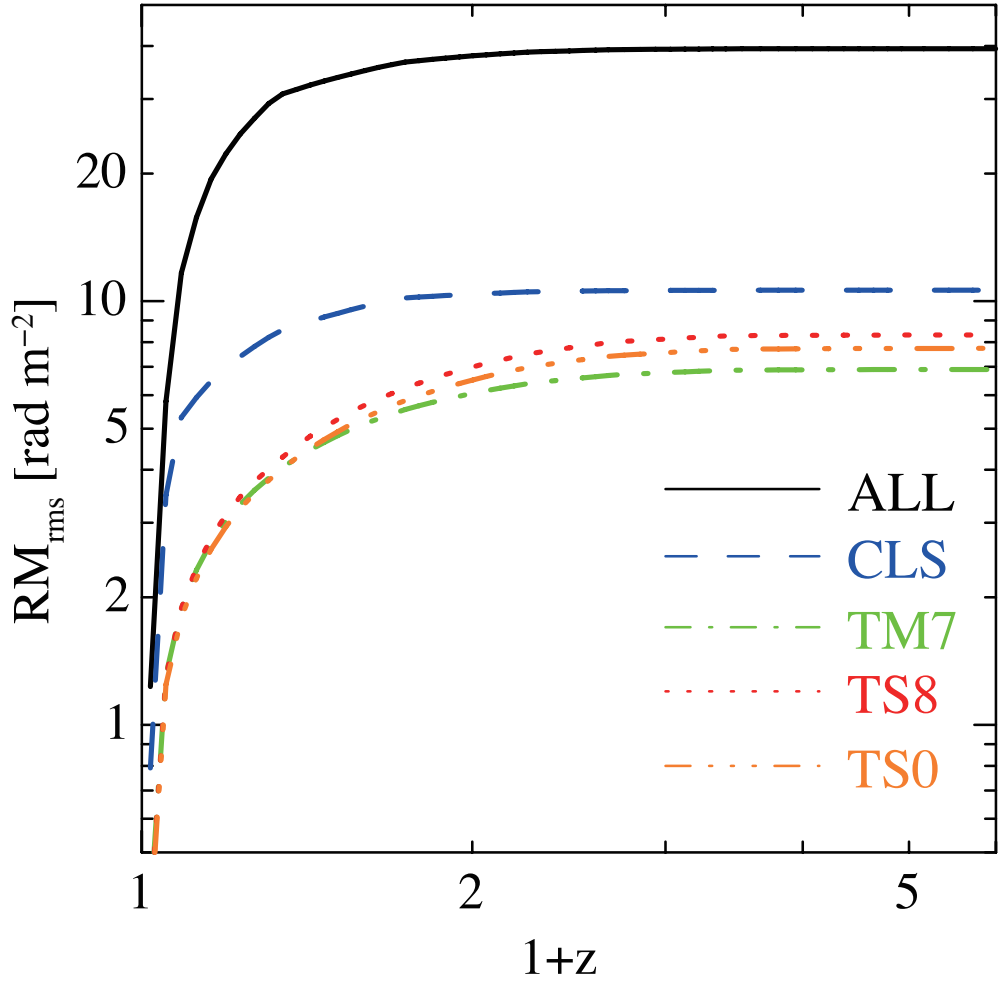}
\caption{${\rm RM}_{\rm rms}$, the rms value of RM, integrated up to the redshift depth $z=5$, for ALL, CLS, TM7, TS8, and TS0 models. The average for $2048^2 \times 200$ pixels is shown.\label{f5}}
\end{figure}

Figure \ref{f5} shows ${\rm RM}_{\rm rms}$, the rms value of RM, as a function of the redshift depth for ALL and all cluster subtraction models. ${\rm RM}_{\rm rms}$ increases steeply for $z \la 1$, and saturates for $z \ga 1$. The saturated value of ${\rm RM}_{\rm rms}$ is $\sim 10$ ${\rm rad~m^{-2}}$ for CLS and $\sim 7 - 8\ {\rm rad~m^{-2}}$ for TM7, TS8, and TS0, while it is $\sim 40$ ${\rm rad~m^{-2}}$ for ALL. Again the high value for ALL is attributed to clusters and groups. So if we observe the RM through the cosmic web outside of clusters, we would get $\sim 7 - 8\ {\rm rad~m^{-2}}$ for ${\rm RM}_{\rm rms}$.

We note that the saturated value depends on the redshift distribution of radio sources, but only weakly. In our model the redshift distribution peaks at $z \la 1$ (see Figure \ref{f1}). We carried out comparison runs, in which we put all radio sources at $z=5$. Then, the saturated value of ${\rm RM}_{\rm rms}$ increased slightly; specifically, for TM7, the value increased from $\sim 8$ to $\sim 9$ ${\rm rad~m^{-2}}$.

\begin{figure}[t]
\figurenum{6}
\epsscale{0.9}
\plotone{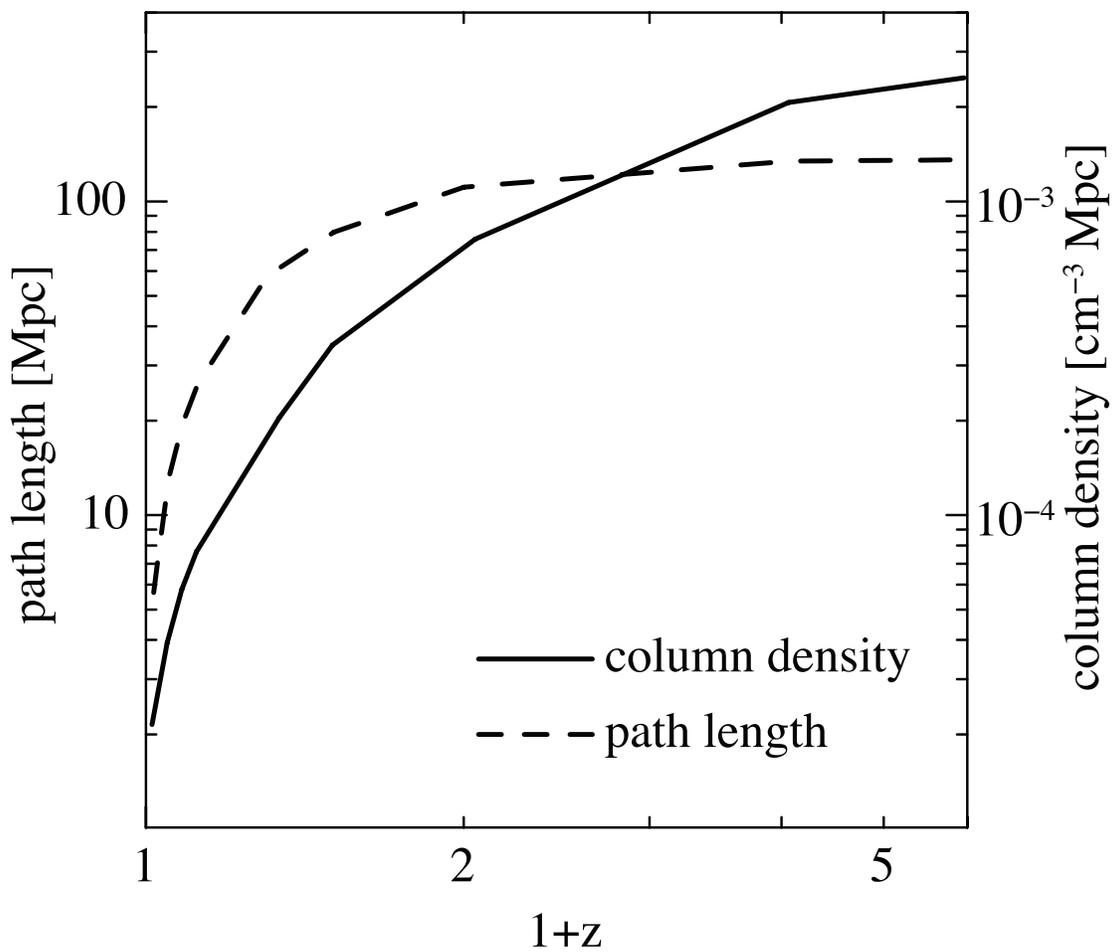}
\caption{
Path length (dashed) and column density (solid) across the WHIM with the temperature of $10^5$ K $<T<$ $10^7$ K, as a function of the redshift depth. The average for $2048^2 \times 200$ pixels is shown.\label{f6}}
\end{figure}

The saturation for $z \ga 1$ is mostly due to the convergence of the path length at $z \sim 1$ and the $(1+z)^{-2}$ dependence of RM (see Equation (\ref{eq5})). To see this, we calculated the path length and also the column density across the WHIM with $10^5$ K $<T<$ $10^7$ K as a function of the redshift depth. Figure \ref{f6} shows the resulting path length and column density. The path length saturates at $\sim$ 100 Mpc for $z \ga 1$. On the other hand, up to $z=1$, the column density reaches $\sim 40\ \%$ of the saturated value. But the contribution to RM from higher redshift should diminish due to the $(1+z)^{-2}$ dependence.

We can roughly estimate the saturated value of ${\rm RM}_{\rm rms}$, as follows. Since the inducement of RM is a random walk process, we may take the value as $\sim {\rm RM}_{\rm rms,\ filament} \times (N_{\rm filaments})^{\sfrac1/2}$, where ${\rm RM}_{\rm rms,\ filament}$ is the rms RM induced by a single filament and $N_{\rm filaments}$ is the number of encounters of filaments. AR10 estimated that ${\rm RM}_{\rm rms,\ filament} \sim 1.5\ {\rm rad~m^{-2}}$. Since the typical width of filaments is several $h^{-1}$ Mpc \citep[see, e.g.,][]{rkhj03,krcs05}, the path length of $\sim$ 100 Mpc corresponds to $N_{\rm filament} \sim 25$. Then, we get ${\rm RM}_{\rm rms} \sim 1.5 \times \sqrt{25} \sim 7.5$ ${\rm rad~m^{-2}}$, which agrees well with those for TM7, TS8, and TS0.

\subsection{Power Spectrum and Structure Function}
\label{s3.4}

\begin{figure}[t]
\figurenum{7}
\epsscale{0.8}
\plotone{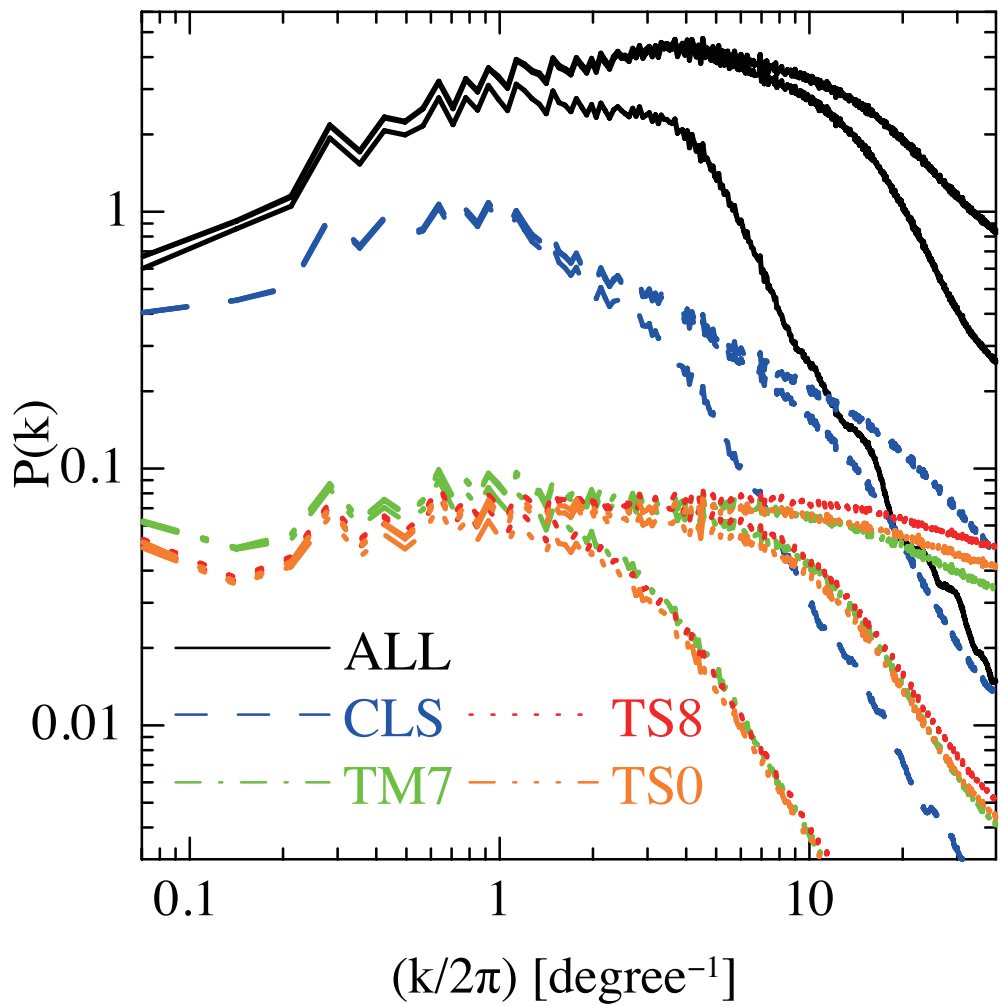}
\caption{Two-dimensional power spectrum of RM for ALL, CLS, TM7, TS8, and TS0 models. The average for 200 stacking simulations is shown. For each model, the power spectra integrated up to the redshift depth z= 0.05, 0.3, and 5.0 are shown.\label{f7}}
\end{figure}

We also calculated the two-dimensional power spectrum with the sky map of RM. Figure \ref{f7} shows the resulting spectrum, averaged over those from 200 maps, for redshift depths $z=0.05$, 0.3, and 5. For our best models TS8 and TS0, at the depth of $z=0.05$, the power spectrum peaks at $\sim 1^\circ$, which corresponds to the proper length of $\sim 2.5\ h^{-1}$ Mpc at that redshift. As the redshift depth increases, the power adds up mostly at smaller scales. At $z=5$, the power spectrum shows a broad plateau over $\sim 1 - 0.1^\circ$ with peak around $\sim 0.15^\circ$. The peak angle corresponds to the proper length of $\sim 3\ h^{-1}$ Mpc at $z=1$ and $\sim 2.5\ h^{-1}$ Mpc at $z=5$. With the typical radius of filaments of a few $h^{-1}$ Mpc, we argue that the peak in the power spectrum reflects the scale of filaments.

The point that the power spectrum peaks at a small angular scale of $\sim 0.15^\circ$ could be the key to the search of the RM due to the IGMF in future RM surveys such as the SKA survey and the ASKAP POSSUM; it can be used to remove the galactic foreground. Towards the galactic poles, the galactic foreground is expected to have the peak of the power spectrum at a much larger scale of $\sim 10^\circ$ or so \citep[e.g.,][]{fsss01}. Then, it would be possible to separate the extragalactic component from the galactic foreground in the Fourier space, when RM surveys become available. The issue of the galactic foreground, along with other uncertainties, in the study of the extragalactic component of RM will be discussed in separate papers.

From the observer's point of view, the structure function (SF) is a statistical quantity which is easier to be obtained. The the $n$-th order SF is defined as 
\begin{equation}\label{eq12}
S_n(r)=\langle|RM(\vec{x}+\vec{r})-RM(\vec{x})|^n\rangle _{\vec{x}}
\end{equation}
with $r=|\vec{r}|$, where the subscript indicates the averaging over the data domain. It can be easily calculated for the data of arbitrary domain with irregular sampling intervals. Figure \ref{f8} shows the second-order SF $(n=2)$ for the depth of $z=5$ in our model, again averaged over those from 200 maps. The SF increases slightly at small angular separations, $\la 0.2^\circ$, and saturates and stays flat at larger separations. The second-order SF is basically the Fourier transform of the power spectrum. The saturation at $\sim 0.2^\circ$ reflects the fact that the power spectrum has the peak around that angular scale, which is indeed the case as shown in Figure \ref{f7}.

Recently, for instance, \citet{mao10} and \citet{sts11} reported the second-order SF from RM surveys towards the galactic poles; their data are also shown in Figure \ref{f8}. The SF towards the south pole is relatively flat on the angular scale of $\sim 1 - 10^\circ$, while it decreases with decreasing angular separation at smaller scales. On the other hand, the SF towards the north pole decreases with decreasing angular separation on the scale of $\sim 1 - 10^\circ$, while it is very noisy but looks flat at smaller scales (the noise is partly attributed to the Coma cluster in the field). The SF towards the south pole is somewhat larger than that toward the north pole. As \citet{sch10} and \citet{sts11} noted, however, the error in the observed SFs is still too large to allow any concrete conclusion. Nevertheless, our results suggest that a substantial fraction of the RM towards the galactic poles may be attributed to the IGMF. As a matter of fact, our estimation of the extragalactic contribution to RM, $\sim 7 - 8\ {\rm rad~m^{-2}}$, agrees with the estimation of $\sim 6\ {\rm rad~m^{-2}}$ from a survey by \citet{sch10}.

\section{Summary}
\label{s4}

Measuring Faraday RM is one of a few available methods to investigate the IGMF in the LSS of the universe, especially in filaments, which still remains largely unknown. Using a model IGMF derived from cosmological structure formation simulations, we examined the RM through the cosmic web outside of clusters, by stacking simulation data up to redshift $z=5$ and taking account of the redshift distribution of radio sources. We found that the magnitude of the RM increases with the path length; the path length and so the value of ${\rm RM}_{\rm rms}$ increases with the redshift depth of radio sources for $z \la 1$ and saturates for $z \ga 1$. The saturated value is predicted to be ${\rm RM}_{\rm rms} \sim 7 - 8 \ {\rm rad~m^{-2}}$. We also found that the PDF of $|{\rm RM}|$ follows the lognormal distribution. The power spectrum of the RM shows a broad plateau over the angular scale of $\sim 1 - 0.1^\circ$ and peaks at $\sim 0.15^\circ$. The second-order SF shows a flat profile in the angular separation of $\ga 0.2^\circ$.

Our results could provide useful insights for planning RM surveys with the SKA, and upcoming SKA pathfinders, ASKAP, MEERKAT, and LOFAR. Also we point that our results could be tested with such surveys. For instance, our power spectrum of RM peaks at the scale of $\sim 0.15^\circ$. A very fine RM map with the average separation of sources $\sim 1$ arcmin, which is expected to be achieved with the SKA, should have enough angular resolution to see the power spectrum peak we predict.

\acknowledgments

\begin{figure}[t]
\figurenum{8}
\epsscale{0.8}
\plotone{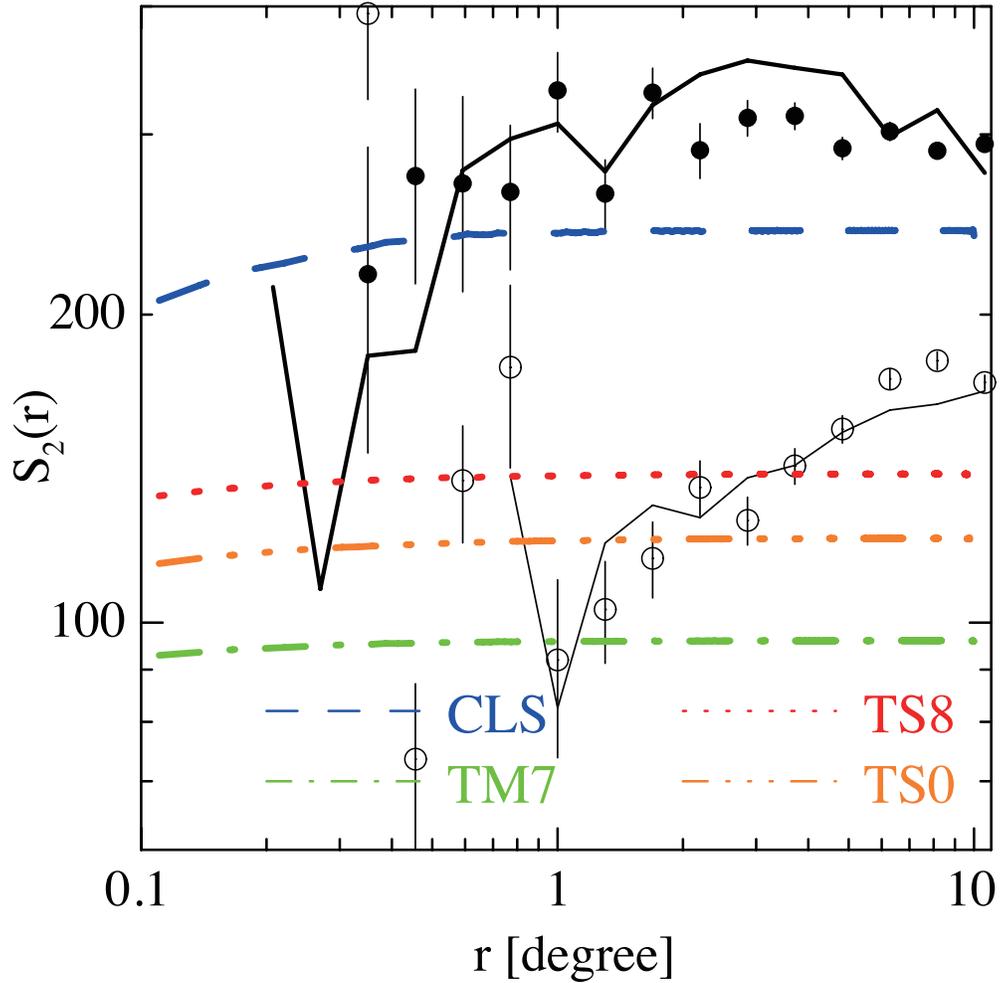}
\caption{Two-dimensional, second-order structure function of RM for CLS, TM7, TS8, and TS0 models. The average for 200 stacking simulations is shown. Also shown are the second-order structure functions from \citet{mao10} (circles) and \citet{sts11} (solid lines). Filled circles and thick lines are toward the south galactic pole, and open circles and thin lines are toward the north galactic pole. They are the same as those in Figure 14 of \citet{sts11}.\label{f8}}
\end{figure}

We thank J. M. Stil for proving the data used in Figure \ref{f8} and H. Kang for comments on the manuscript. The work was supported by the National Research Foundation of Korea through grant 2007-0093860.

\clearpage

\clearpage

\clearpage

\clearpage

\clearpage

\clearpage

\clearpage

\clearpage


\begin{thebibliography}{}
\bibitem[Akahori \& Ryu(2010)]{ar10}
	Akahori, T., \& Ryu, D. 2010, \apj, 723, 476 (AR10)
\bibitem[Ando et al.(2010)]{ads10}
	Ando, M., Doi, K., \& Susa, H. 2010, \apj, 716, 1566
\bibitem[Beck(2009)]{beck09}
	Beck, R. 2009, \rmxaa\ (Serie de Conferencias), 36, 1
\bibitem[Biermann(1950)]{bie50}
	Biermann, L. 1950, Z. Naturforsch, A, 5, 65
\bibitem[Bonafede et al.(2010)]{bfmg10}
	Bonafede, A., Feretti, L., Murgia, M., Govoni, F., Giovannini, G.,
	Dallacasa, D., Dolag, K., \& Taylor, G. B.  2010, \aap, 513, 30
\bibitem[Bond et al.(1996)]{bkp96}
    Bond, J. R., Kofman, L., \& Pogosyan, D. 1996, \nat, 380, 603
\bibitem[Carilli \& Taylor(2002)]{ct02}
	Carilli, C. L. \& Taylor, G. B. 2002, \araa, 40, 319
\bibitem[Carilli \& Rawlings(2004)]{cr04}
	Carilli, C. L. \& Rawlings S. 2004, New A Rev., 48, 1
\bibitem[Cen \& Ostriker(1999)]{co99}
	Cen, R. \& Ostriker, J. P. 1999, \apj, 514, 1
\bibitem[Cho et al.(2009)]{cvblr09}
	Cho, J., Vishniac, E. T., Beresnyak, A., Lazarian, A., \& Ryu, D. 2009, \apj, 693, 1449
\bibitem[Cho \& Ryu(2009)]{cr09}
	Cho, J., \& Ryu, D. 2009, \apjl, 705, L90 (CR09)
\bibitem[Clarke et al.(2001)]{ckb01}
	Clarke, T. E., Kronberg, P. P., \& B\"{e}hringer, H. 2001, \apjl, 547, L111
\bibitem[Clarke(2004)]{cla04}
	Clarke, T. E. 2004, J. Korean Astron. Soc. 37, 337
\bibitem[da Silva et al.(2000)]{sblt00}
	da Silva, A. C., Barbosa, D., Liddle, A. R., \& Thomas, P. A. 2000, \mnras, 317, 37
\bibitem[Das et al.(2008)]{dkrc08}
    Das, S., Kang, H., Ryu, D., \& Cho, J. 2008, \apj, 682, 29
\bibitem[Dolag \& Stasyszyn(2009)]{ds09}
	Dolag, K., \& Stasyszyn, F. 2009, MNRAS, 398, 1678
\bibitem[Donnert et al.(2009)]{ddlm09}
    Donnert, J., Dolag, K., Lesch, H., \& M\"{u}ller, E. 2009, \mnras, 392, 1008
\bibitem[Dubois \& Teyssier(2008)]{dub08}
	Dubois, Y. \& Teyssier, R. 2008, \aap, 482, L13
\bibitem[Frick et al.(2001)]{fsss01}
	Frick, P., Stepanov, R., Shikurov, A., \& Sokoloff, D. 2001, \mnras, 325, 649
\bibitem[Gaensler et al.(2010)]{glt10}
	Gaensler, B., Landecker, T., \& Taylor, R. 2010, Bulletin of the American Astronomical Society, 42, 515
\bibitem[Gnedin et al.(2000)]{gfz00}
	Gnedin, N. Y., Ferrara, A., \& Zweibel, E. G. 2000, \apj, 539, 505
\bibitem[Govoni et al.(2010)]{gov10}
	Govoni, F., Dolag, K., Murgia, M., Feretti, L., Schindler, S., Giovannini, G., Boschin, W., Vacca, V., \& Bonafede, A. 2010, \aap, 522, 105
\bibitem[Guidetti et al.(2008)]{gmgp08}
	Guidetti, D., Murgia, M., Govoni, F., Parma, P., Gregorini, L., deRuiter, H. R., Cameron, R. A., \& Fanti, R. 2008, \aap, 483, 699
\bibitem[Hoeft et al.(2008)]{hoeft08} 
    Hoeft, M., Bruggen, M., Yepes, G., Gottlober, S., \& Schwope, A., 2008, \mnras, 391, 1511
\bibitem[Hoshino et al.(2010)]{hos10}
	Hoshino, A., et al. 2010, \pasj, 62, 371
\bibitem[Kang et al.(2007)]{krco07}
    Kang, H., Ryu, D., Cen, R., \& Ostriker, J. P. 2007, \apj, 669, 729
\bibitem[Kang et al.(2005)]{krcs05}
    Kang, H., Ryu, D., Cen, R., \& Song, D. 2005, \apj, 620, 21
\bibitem[Krause et al.(2009)]{kra09}
	Krause, M., Alexander, P., Bolton, R., Geisb\"{u}sch, J., Green, D. A., \& Riley, J. 2009, \mnras, 400, 646
\bibitem[Kulsrud et al.(1997)]{kcor97}
	Kulsrud, R. M., Cen, R., Ostriker, J. P., \& Ryu, D. 1997, \apj, 480, 481
\bibitem[Langer et al.(2005)]{lap05}
	Langer, M., Aghanim, N., \& Puget, J. L. 2005, \aap, 443,367
\bibitem[Mao et al.(2010)]{mao10}
	Mao, S. A., Gaensler, B. M., Haverkorn, M., Zweibel, E. G., Madsen, G. J., McClure-Griffiths, N. M., Shukurov, A. \& Kronberg, P.~P. 2010, \apj, 714, 1170
\bibitem[Miniati \& Bell(2011)]{mb11}
    Miniati, F. \& Bell, A. R. 2011, \apj, 729, 73
\bibitem[Peebles(1993)]{peeb93}
    Peebles, P. J. E. 1993, Principles of Physical Cosmology (Princeton: Princeton Univ. Press)
\bibitem[Pfrommer et al.(2006)]{psej06}
    Pfrommer, C., Springel, V., En{\ss}lin, T. A., \& Jubelgas, M. 2006, \mnras, 367, 113
\bibitem[Rybicki \& Lightman(1979)]{rl79}
	Rybicki, G. B., \& Lightman, A. P. 1979, Radiative Processes in Astrophysics (New York: Wiley-Interscience)
\bibitem[Rasmussen \& Pedersen(2001)]{rp01}
	Rasmussen, J., \& Pedersen, K. 2001, \apj, 559, 892
\bibitem[Ryu et al.(1993)]{rokc93}
    Ryu, D., Ostriker, J. P., Kang, H., \& Cen, R. 1993, \apj, 414, 1
\bibitem[Ryu et al.(1998)]{rkb98}
	Ryu, D., Kang, H., \& Biermann, P. L. 1998, \aap, 335, 19
\bibitem[Ryu et al.(2003)]{rkhj03}
	Ryu, D., Kang, H., Hallman, E., \& Jones, T. W. 2003, \apj, 593, 599
\bibitem[Ryu et al.(2008)]{rkcd08}
	Ryu, D., Kang, H., Cho, J., \& Das, S. 2008, Science, 320, 909 (R08)
\bibitem[Ryu et al.(2010)]{rstt10}
	Ryu, D., Schleicher, D. R. G., Treumann, R. A., Tsagas, C. G., \& Widrow, L. M. 2010, Space Sci. Rev., submitted
\bibitem[Schleicher et al.(2010)]{sbsa10}
	Schleicher, D. R. G., Banerjee, R., Sur, S., Arshakian, T. G., Klessen, R. S., Beck, R., \& Spaans, M. 2010, \aap, 522, A115
\bibitem[Schlickeiser \& Shukla(2003)]{ss03}
    Schlickeiser, R., \& Shukla, P. K. 2003, \apjl, 599, L57
\bibitem[Schnitzeler(2010)]{sch10}
	Schnitzeler, D. H. F. M 2010, \mnras, 409, 99
\bibitem[Skillman et al.(2008)]{skillman08}
    Skillman, S. W., O'Shea, B. W., Hallman, E. J., Burns, J. O., \& Norman, M. L. 2008, \apj, 689, 1063
\bibitem[Stasyszyn et al.(2010)]{sndbd10}
	Stasyszyn, F., Nuza, S. E., Dolag, K., Beck, R., \& Donnert, J. 2010, \mnras, 408, 684
\bibitem[Stil et al.(2011)]{sts11}
	Stil, J. M., Taylor, A. R., \& Sunstrum, C. 2011, \apj, 764, 4
\bibitem[Taylor et al.(2009)]{tss09}
	Taylor, A. R., Stil, J. M., \& Sunstrum, C. \apj, 702, 1230
\bibitem[Widrow et al.(2010)]{wrss10}
	Widrow, L. M., Ryu, D., Schleicher, D. R. G., Subramanian, K., Tsagas, C. G., \& Treumann, R. A. 2010, Space Sci. Rev., submitted
\bibitem[Willman et al.(2008)]{will08}
	Willman, R. J., et al. 2008, \mnras, 388, 1335
\bibitem[Xu et al.(2006)]{xkhd06}
	Xu, Y., Kronberg, P. P., Habib, S., \& Dufton, Q. W. 2006, \apj, 637, 19
\bibitem[Xu et al.(2008)]{xu08}
	Xu, H., O'shea, B. W., Collins, D. C., Norman, M. L. Li, H., \& Li, S. 2008, \apjl, 688, L57
\bibitem[Vazza et al.(2009)]{vazza09}
    Vazza, F., Brunetti, G., \& Gheller, C. 2009, \mnras, 395, 1333
\bibitem[Vogt \& En{\ss}lin(2005)]{ve05}
	Vogt, C. \& En{\ss}lin, T. A. 2005, \aap, 434, 67
\end{thebibliography}
\end{document}